This article has been accepted for publication in IEEE Transactions on Instrumentation and Measurement. This is the author's version which has not been fully edited and content may change prior to final publication. Citation information: DOI 10.1109/TIM.2025.3568972

# Hybrid Resolver Model Generalization for Fault Condition Modeling: A Promising Tool for Reliability Study

MohammadSadegh KhajueeZadeh, Farid Tootoonchian, *Senior Member, IEEE*, and Ali Pourghoraba

***Abstract*— Resolvers, like all electromagnetic devices, are constantly under investigation, both operationally and structurally. In this regard, proposing a modeling methodology that can save significant time without compromising accuracy is a big honor, as each of the other analytical methodologies fails to meet at least one of the above conditions. In this study, a generalized hybrid model is suggested that, in addition to the above benefits, has sufficient capability to ease reliability study in the field of resolvers, where a large number of faulty conditions must be investigated under different operating conditions, including changes in angular velocity, voltage, and frequency of excitation—all of which are highlighted in the context of fault coverage. This model also serves as a promising tool for generating large datasets, which is advantageous for fault diagnosis. A resolver with a non-uniform airgap is chosen as a case study to challenge the suggested model, particularly in relation to eccentricity faults. We generalize the suggested model to account for the most common faulty conditions of resolvers: inter-turn short circuits in signal and excitation windings, as well as static and dynamic eccentricity faults. The close agreement between the results of the suggested model and those from Time-Stepping Finite Element Analysis (TS-FEA), along with significant time savings in both healthy and faulty conditions, highlights the generality and proficiency of the suggested model. Finally, the case study is prototyped, and we verify the accuracy of the suggested model experimentally.**

## I. Introduction

GLOBAL warming is the main concern that ignites a wave of interest among scholars to find solutions for at least reducing its growth. Electric cars are one such solution that has begun to gain attention. The heart of these cars is electric motors, which are most often Permanent Magnet Synchronous Motors (PMSMs) due to their well-known advantages, such as high efficiency and dynamics. The control of these motors heavily has a dependency on the data that the drive system accesses through sensors, such as current and rotor angle, among others. In this context, the accuracy of the rotor angle sensor, as shown in [1], has a significant effect on drive stability and even more on the reliability of the entire system.

In this regard, almost all designers in the industry consider resolvers as their first choice, since they have shown satisfactory tolerance in the face of harsh operating conditions, including excessive noise, heat, and dust—significant weaknesses that encoders suffer from due to their sensitive physical configuration [2]. Although resolvers have shown their proficiency, there is still a necessity to study their behavior under different faulty conditions, particularly in terms of lifetime and fault management. In lifetime management [3], we analyze and estimate the lifetime of the investigated system; the resolver is a subsystem within the larger electric car system. Following this, we make efforts to redesign resolvers to achieve better accuracy and fault tolerance, synonymous with a longer lifetime for the entire system. On the other hand, in fault management [4], we will suggest different methodologies to engineers that can be advantageous for fault diagnosis, isolation, and even mitigation in resolvers. A common theme among all these is the examination of resolvers when one or overlapping faulty conditions occur; therefore, a designer should have a clear understanding of what will occur to the resolver under faulty conditions. Prototyping a resolver, even in healthy conditions, is time-consuming and costly; thus, if a designer wants to study a resolver's reliability (each of lifetime and fault management), fabricating a large number of faulty resolvers is not pragmatic.

In this regard, different modeling methodologies have been suggested to make this faster and more cost-effective. However, we should search for a candidate that can offer all metrics together: accuracy and time efficiency. Hence, Time-Stepping Finite Element Analysis (TS-FEA) is not the best choice [5]; although it gives high accuracy in modeling, each run takes a significant amount of time due to the high-frequency excitation of the resolver, and subsequently, the high sampling frequency, requiring for signal voltage calculations to achieve better results from signal processing. If we are looking to save time, analytical models are better choices. In this context, Magnetic Equivalent Circuit (MEC) and Winding Function (WF) are the analytical modeling methodologies that have been employed for fault modeling in resolvers [6]-[10]. It is evident that each of them struggles with disadvantages, particularly when dealing with diverse physical configurations and faulty conditions, impacting both time efficiency and accuracy.

In [6]-[8], the MEC methodology forms the basis for

M.S. KhajueeZadeh is with Electrical and Computer Engineering Department, The University of British Columbia, Canada, (e-mail: mohammadkhajuee@ece.ubc.ca). F. Tootoonchian and A. Pourghoraba are with the Department of Electrical Engineering, Iran University of Science and Technology, Tehran 16846-13114, Iran (e-mail: tootoonchian@iust.ac.ir; apourghoraba@gmail.com).

modeling faulty conditions, such as inter-turn short circuit or eccentricity. In [6], inter-turn short circuit and eccentricity were investigated for a resolver with two stator winding configurations: non-overlapping and overlapping windings. This investigation came after earlier work in [7] and [8], including Permanent Magnet (PM) resolvers with non-overlapping stator windings and three rotor contours in [7], and resolvers with three stator winding configurations and four rotor contours in [8]. In [6]-[8], different configurations were investigated, representing changes in geometries. Each configuration has a new stator and rotor magnetic circuit network, reducing the generality of MEC modeling and introducing extra complexity and time burdens for modeling. Moreover, the MEC methodology faces challenges in predicting flux tubes in airgap regions, particularly regarding leakages and slot skewing [9]. While a 3D MEC can address these weaknesses, the time factor—decisive in reliability examinations— must be regarded, as a large number of faulty conditions should be investigated to ensure resolver tolerability. Additionally, the accuracy of MEC and the complexity of magnetic circuit networks have a direct dependency. Therefore, modeling faulty conditions with accuracy close to TS-FEA—especially in geometries with high complexity—may render MEC less effective for saving time. In [10] and [11], the WF methodology serves as the basis for modeling faulty conditions of resolvers. WF begins with calculating the magnetic flux intensity, subsequently magnetic flux density and flux linkages. In eccentricity modeling we should modify the airgap function according to the type—static or dynamic—leading to changes in magnetic flux intensity [11]. For inter-turn short circuit modeling, changes in winding turn numbers and short-circuit resistance must be regarded when writing Winding's Turn Functions (WTFs) [10]. In [10] and [11], the Fast Fourier Transform (FFT) of WTFs was shown, and the most effective terms were chosen, simplifying the model and the calculation of magnetic flux density and flux linkages. However, faulty conditions such as eccentricity excite other terms, and ignoring them misses decisive changes in the motor and drive system. WF also struggles with modeling non-uniform airgaps and slotting [12]. Other analytical methodologies, such as Field Reconstruction Method (FRM) [9] and Subdomains Model (SDM) [13], face challenges with irregular designs and fault conditions. Precomputing magnetic field solutions in FRM limits accuracy in geometries with high complexity, requiring readjustments when designs change. Moreover, SDM struggles with eccentricity faults due to its dependency on a uniform airgap.

In [14], we suggested the theory of the hybrid resolver model for the first time, which has sufficient generalization and capability in modeling different physical configurations [15], winding configurations, and faulty conditions of resolvers, both electrically and mechanically. This model serves as a stepping stone in the reliability study of resolvers [2], saving significant time—especially when dealing with a large number of analyses and iterative designs [16]—while maintaining high accuracy similar to TS-FEA. Additionally, it benefits us by providing an online link between the primary motor, resolver, load, and control algorithm as an integrated drive set [2]. However, this study aims to further generalize the hybrid resolver model from [14], specifically improving the time efficiency of modeling different winding configurations and faulty conditions without significantly compromising accuracy. The organization of this study is as follows: first, we clarify the background theory of the generalized hybrid model, referring to the basic idea in [14]. Next, how faulty conditions are integrated into the suggested model, specifically for inter-turn short circuit and eccentricity categories, is written in detail. Section IV focuses on the proficiency and generality of the suggested model in handling different winding configurations and faults, along with an evaluation of the model's accuracy against TS-FEA is given. In Section V, the case study prototype undergoes fault injection to verify the suggested model experimentally. Finally, the results and conclusions are given briefly. The key contributions of this study are highlighted below: (1) A generalized semi-analytical modeling method for diverse configurations, winding designs, and fault conditions. (2) Major gains in time efficiency and accuracy—at least 96% faster than TS-FEA with under 13% worst-case accuracy loss. (3) Integration of resolver faults (e.g., inter-turn short circuits, eccentricity) and new winding designs through a one-time preparatory stage, without requiring the re-execution of TS-FEA.

## II. Background Theory of the Hybrid Model

Given that resolvers work in the linear and bottom zones of the B-H curve of their ferromagnetic cores, which is synonymous with the linear behavior of resolvers, we can show their electromagnetic behavior as follows [14]:

$$\begin{bmatrix}\lambda_e\\ \lambda_s\\ \lambda_c\end{bmatrix}=\begin{bmatrix}L_{ee}(\theta) & L_{es}(\theta) & L_{ec}(\theta)\\ L_{se}(\theta) & L_{ss}(\theta) & L_{sc}(\theta)\\ L_{ce}(\theta) & L_{cs}(\theta) & L_{cc}(\theta)\end{bmatrix}\begin{bmatrix}i_e\\ i_s\\ i_c\end{bmatrix} \quad (1)$$

Here, $e$, $s$, and $c$ symbolize the excitation, sine, and cosine windings, respectively. Moreover, $\lambda_\gamma$, $i_\gamma$, and $L_{\gamma\gamma}$ denote the flux linkage, current, and self-inductance of each winding, respectively, while $L_\gamma\xi$ denotes the mutual inductance between two windings, and $\theta$ is the symbol of rotor angle. In this context, the role of generating the magnetic field rests solely on the excitation winding, as the currents in the sine and cosine signal windings are negligible. In this regard, we can simplify (1) as written in (2).

$$[\lambda_e \quad \lambda_s \quad \lambda_c]^T=[L_{ee}(\theta) \quad L_{se}(\theta) \quad L_{ce}(\theta)]^T i_e \quad (2)$$

We also refer to the fact that the resolver works periodically and has the capability to write the terms of $[L_{ee}(\theta) \quad L_{se}(\theta) \quad L_{ce}(\theta)]^T$ using Fourier series, as shown below [14]:

$$\begin{aligned}L_{ee}(\theta)&=L_{e0}+\sum l_{en}\sin(n[\theta-\theta_0]+\varphi_{en})\\ L_{se}(\theta)&=L_{s0}+\sum L_{sn}\sin(n[\theta-\theta_0]+\varphi_{sn})\\ L_{ce}(\theta)&=L_{c0}+\sum L_{cn}\sin(n[\theta-\theta_0]+\varphi_{cn})\end{aligned} \quad (3)$$

Here, $l_{\gamma n}$ and $\varphi_{en}$ are the Fourier series coefficients and angles, respectively, while $\theta_0$ denotes the rotor angle at time = 0 s. Ideally, there is no load on the resolver shaft, and it rotates synchronously with the primary motor; therefore, $\theta = \theta_0 + \int \omega dt$ is true. The resolver's excitation is a high-frequency voltage; therefore, since the excitation frequency should be much higher than the rotation frequency, $L_{ee}(\theta)$ is approximately constant. In this regard, the angular velocity voltage is negligible $(dL_{ee}(\theta)/d\theta = 0)$ [15].

$$v_e = R_e i_e + d\lambda_e/dt \quad (4)$$

Here, $v_e$, the excitation voltage, can be written as $v_m \cos \omega_e t$, where $v_m$ is the voltage magnitude, and $f_e$ in $\omega_e = 2\pi f_e$ denotes the excitation frequency. In (4), $R_e$ denotes the primary resistance of the excitation winding. As in a transformer, supplying the primary side generates sine and cosine signal voltages ($v_s$ and $v_c$) on the secondary windings:

$$v_s = d\lambda_s/dt, \qquad v_c = d\lambda_c/dt \quad (5)$$

Processing these signal voltages yields their arctangent, resulting in the resolver rotor angle, $\theta = \tan^{-1}(v_s/v_c)$. A discrepancy always exists between the primary motor's rotor angle and that of the resolver; this is known as the rotor angle error, which is a waveform. To gain a better understanding of resolver accuracy, scholars define two metrics: the Average of Absolute Position Error (AAPE) and the Maximum Position Error (MPE). According to (2), we can infer that the terms of $[L_{ee}(\theta) \quad L_{se}(\theta) \quad L_{ce}(\theta)]^T$ are the basis for resolver modeling using hybrid model theory. In [14], TS-FEA was the main tool for computing these terms. In this regard, we should design the resolver in FEM software; then, by supplying the excitation winding with a constant current, such as 10 mA, we will obtain $[L_{ee}(\theta) \quad L_{se}(\theta) \quad L_{ce}(\theta)]^T$ for a rotation from 0 to 360 degrees. What is highlighted in the methodology of [14] is that we should apply any changes in winding configuration, such as making decisions about the number of turns for each coil within each slot, the winding diagram, and even the faulty conditions that we want to study, before running TS-FEA. Regardless of these, the suggested hybrid model in [14] is much faster than TS-FEA in resolver modeling, with accuracy close to that of TS-FEA. However, what we are looking for is a modeling methodology that makes it easy for us to study a large number of faulty conditions for a resolver.

In the generalization of the hybrid model, we also have a designed resolver as a case study, with an unchangeable number of slots and saliency for the stator or rotor; there are no constraints regarding the configuration, similar to what was shown in [14] and [15]. On the contrary, unlike [14], we do not apply the final number of turns or winding diagram, nor do we inject any faults; rather, we will calculate the terms of $[L_{ee}(\theta) \quad L_{se}(\theta) \quad L_{ce}(\theta)]^T$ for a resolver that has single-turn coils within all of its slots. Previously, in [14], the signal winding was categorized into sine and cosine windings, and we should calculate all the terms in $[L_{ee}(\theta) \quad L_{se}(\theta) \quad L_{ce}(\theta)]^T$ completely. However, now we can simply calculate $L_{signal,e}(\theta)$ and $L_{ee}(\theta)$, which align with the single-turn coils in the stator or rotor slots. After performing the TS-FEA analysis, we can easily derive $L_{se}(\theta)$ and $L_{ce}(\theta)$ for $L_{signal,e}(\theta)$ separately by applying the winding configuration that we want to investigate. Accordingly, the generalized hybrid model is even faster at initializing. The result will be an $N \times N$ matrix for $L_{ee}(\theta)$ and an $N \times M$ matrix for $L_{signal,e}(\theta)$. $N$ and $M$ denote the number of slots/teeth in the excitation and signal windings. Now we are free to define different mixtures of winding configurations and turn numbers for each coil within the slots.

If $T_{si}$ is the number of turns at the $i$-th tooth of the sine winding, and $T_{ej}$ is the number of turns at the $j$-th tooth of the excitation winding, we can calculate $L_{se}$ between the sine and excitation windings as written in (6-1) by summing all terms of matrix $M_{se}$ in (6-2), when the coils are in series. For each of $T_{si}$ and $T_{ej}$ in $M_{se}$, we should take into account their magnitude and direction. Positive and negative signs handle the current direction, while zero means there is no coil on that tooth.

$$L_{se} = \sum_{i=1}^{M} \sum_{j=1}^{N} M_{se}[i,j] \quad (6\text{-}1)$$

$$M_{se} = \begin{bmatrix} T_{s_1}T_{e_1}L_{s_1e_1} & T_{s_1}T_{e_2}L_{s_1e_2} & \cdots & T_{s_1}T_{e_N}L_{s_1e_N} \\ T_{s_2}T_{e_1}L_{s_2e_1} & T_{s_2}T_{e_2}L_{s_2e_2} & \cdots & T_{s_2}T_{e_N}L_{s_2e_N} \\ \vdots & \vdots & \ddots & \vdots \\ T_{s_M}T_{e_1}L_{s_Me_1} & T_{s_M}T_{e_2}L_{s_Me_2} & \cdots & T_{s_M}T_{e_N}L_{s_Me_N} \end{bmatrix} \quad (6\text{-}2)$$

In this regard, the methodology is the same for the $L_{ce}$ calculation, as written in (7).

$$L_{ce} = \sum_{i=1}^{M} \sum_{j=1}^{N} M_{ce}[i,j] \quad (7\text{-}1)$$

$$M_{ce} = \begin{bmatrix} T_{c_1}T_{e_1}L_{c_1e_1} & T_{c_1}T_{e_2}L_{c_1e_2} & \cdots & T_{c_1}T_{e_N}L_{c_1e_N} \\ T_{c_2}T_{e_1}L_{c_2e_1} & T_{c_2}T_{e_2}L_{c_2e_2} & \cdots & T_{c_2}T_{e_N}L_{c_2e_N} \\ \vdots & \vdots & \ddots & \vdots \\ T_{c_M}T_{e_1}L_{c_Me_1} & T_{c_M}T_{e_2}L_{c_Me_2} & \cdots & T_{c_M}T_{e_N}L_{c_Me_N} \end{bmatrix} \quad (7\text{-}2)$$

For $L_{ee}$, the calculation is as shown in (8), where $T_i$ and $T_j$ are the number of turns for each excitation winding coil at the $i$-th and $j$-th teeth, respectively. It is worth mentioning that in the first term of (8-1), $l_{e_ie_j}$ in (8-2) is symmetric ($L_{eiej} = L_{ejei}$), and we use a factor of 2 in this context, where $i \neq j$. Ignoring this factor of 2 leads to a discrepancy between the signal voltage results from the hybrid model and TS-FEA. For further details on (8-1) and the role of the factor 2, see the APPENDIX link.

$$L_{ee} = 2\sum_{i=1}^{N} \sum_{j=i+1}^{N} T_iT_jL_{e_ie_j} + \sum_{i=1}^{N} T_i{}^2L_{e_ie_i} \quad (8\text{-}1)$$

$$l_{e_ie_j} = \begin{bmatrix} 0 & T_1T_2L_{e_1e_2} & \cdots & T_1T_NL_{e_1e_N} \\ T_2T_1L_{e_2e_1} & 0 & \cdots & T_2T_NL_{e_2e_N} \\ \vdots & \vdots & \ddots & \vdots \\ T_NT_1L_{e_Ne_1} & T_NT_2L_{e_Ne_2} & \cdots & 0 \end{bmatrix} \quad (8\text{-}2)$$

$$l_{e_ie_i} = diag\,[T_1^2L_{e_1e_1} \;\; T_2^2L_{e_2e_2} \;\; \cdots \;\; T_N^2L_{e_Ne_N}]^T \quad (8\text{-}3)$$

Note that the sampling frequency and meshing quality in TS-FEM directly affect modeling accuracy. Regarding sampling frequency, the Nyquist frequency must be taken into account; the chosen sampling frequency should contain a sufficiently wide frequency ranges to accurately rebuild the original waveforms. This ensures that the terms of $[L_{ee}(\theta) \quad L_{se}(\theta) \quad L_{ce}(\theta)]^T$ can be accurately written as a Fourier series, as shown in (3). According to the Fourier series constraints, only dominant orders are chosen—11 orders for $L_{se}(\theta)$ and $L_{ce}(\theta)$, and 4 for $L_{ee}(\theta)$ due to its near-constant behavior—with adjustments made for high-frequency faults to ensure accuracy. Following (3), we can calculate flux linkages from (2) and subsequently the signal voltages from (5). The Trapezoidal rule stands as the main methodology for solving (4) and (5). Thus far, we have suggested the generalized hybrid resolver model for fault-free operating conditions.

## III. Modeling Faulty Conditions in the Hybrid Model

What we delve into in the following is proposing how faulty conditions, including short circuit and eccentricity (both static and dynamic), can be integrated into the generalized hybrid model without repeatedly using TS-FEA.

### A. Inter-Turn Short Circuits in Signal and Excitation Windings

Taking advantage of the flexibility in applying coil turn numbers and winding diagrams, we can modify (6)-(8) to model inter-turn short circuit faults in signal and excitation windings. Given that signal winding currents are negligible, an inter-turn short circuit within them is synonymous with subtracting the number of turns under the short circuit at the $i$-th tooth ($T_{\text{short},s,i}$ for sine winding and $T_{\text{short},c,i}$ for cosine winding) from the original turn numbers of the coil at that tooth ($T_{si}$ for sine winding and $T_{ci}$ for cosine winding) [10]. Accordingly, in (6) and (7), we use the effective turn numbers for the $i$-th tooth ($T_{eff,s,i}$ and $T_{eff,c,i}$ for sine and cosine windings, respectively) in the calculations, which can be written for signal inter-turn short circuits as follows:

$$\begin{aligned} T_{eff,s,i} &= T_{si} - T_{\text{short},s,i} \\ T_{eff,c,i} &= T_{ci} - T_{\text{short},c,i} \end{aligned} \tag{9}$$

$$\begin{aligned} M_{se}[i,j] &= T_{eff,s,i} \cdot T_{e_j} \cdot L_{s_i e_j} \\ M_{ce}[i,j] &= T_{eff,c,i} \cdot T_{e_j} \cdot L_{c_i e_j} \end{aligned} \tag{10}$$

We can behave similarly when an inter-turn short circuit occurs within the excitation winding, with the subtle difference that the excitation current is not negligible, and the effect of the short-circuit resistance ($R_{sc}$) must be regarded in the effective turn numbers for the $j$-th tooth of the excitation winding ($T_{eff,e,j}$), where $R_e$ is the primary resistance of the excitation winding [6]. In this regard, (6)-(8) can be adjusted as follows:

$$T_{eff,e,j} = T_{ej} - T_{\text{short},e,j} + \left(\frac{R_{sc}}{\frac{R_e}{N}+R_{sc}}\right) T_{\text{short},e,j} \tag{11-1}$$

$$M_{se}[i,j] = T_{s_i} \cdot T_{eff,e,j} \cdot L_{s_i e_j} \tag{11-2}$$

$$M_{ce}[i,j] = T_{c_i} \cdot T_{eff,e,j} \cdot L_{c_i e_j} \tag{11-3}$$

$$\begin{aligned} l_{e_i e_j} &= T_{eff,e,i} \cdot T_{eff,e,j} \cdot L_{e_i e_j} \\ l_{e_i e_i} &= T^2{}_{eff,e,i} \cdot L_{e_i e_i} \end{aligned} \tag{11-4}$$

Here, we consider the resistance of the excitation winding in one tooth as $R_e/N$, assuming that all teeth share the resistance equally in series. According to (11-1), if $R_{sc} \to \infty$, there is no short circuit; conversely, $R_{sc} = 0$ denotes the highest short-circuit intensity [6].

### B. Static and Dynamic Eccentricity Faults

In general, eccentricity, whether static or dynamic, comes with airgap changes that directly affect the terms of $[L_{ee}(\theta) \quad L_{se}(\theta) \quad L_{ce}(\theta)]^T$. Therefore, through different scaling factors, we can integrate eccentricity faults into their calculation. Depending on the resolver's rotor configuration, the airgap can be written as follows:

$$g(\phi) = \frac{G_{min}G_{max}}{(G_{min}+G_{max})+(G_{min}-G_{max})\cos\left(\frac{P}{2}\varphi\right)} \tag{12-1}$$

$$g_0 = (D_s - D_r)/2 \tag{12-2}$$

When the resolver has a rotor with a non-uniform airgap that changes sinusoidally, (12-1) holds, where $\phi$ denotes the angle in cylindrical coordinate of the rotor frame in radians, $G_{min}$ is the minimum airgap length, $G_{max}$ is the maximum airgap length, and $P/2$ denotes the number of the rotor saliency. On the other hand, for wound-rotor resolvers with a rotor of constant outer diameter, (12-2) describes the uniform airgap, where $D_s$ and $D_r$ are the inner diameter of the stator and the outer diameter of the rotor, respectively.

Static eccentricity emerges when the location of the stator's symmetric axis shifts, while the rotor's location and its rotation symmetric axis remain unchanged [2]. In other words, under static eccentricity, the rotor and stator remain in a static configuration, and the rotor does not rotate relative to the stator. If static eccentricity occurs in a resolver with a magnitude of $e_s$ in millimeters and an angular direction of $\theta_{ecc}$ in radians (counterclockwise between 0 and $2\pi$), the scaling factor for static eccentricity, $f_{se}(\phi)$, adjusts the calculations of the terms of $[L_{ee}(\theta) \quad L_{se}(\theta) \quad L_{ce}(\theta)]^T$, as shown in (13-2). The change in the airgap can be written as (13-1) when the resolver undergoes static eccentricity. For simplicity, the angular location of the static eccentricity, $\theta_{ecc}$, can be written as shown in (13-3), specifying the static eccentricity's angular direction relative to the $t$-th tooth of the stator, which remains constant over time.

$$g_{static}(\phi) = g(\phi) + e_s.\cos(\varphi - \theta_{ecc}) \tag{13-1}$$

$$f_{se}(\phi) = g(\phi)/(g(\phi) + e_s.\cos(\varphi - \theta_{ecc})) \tag{13-2}$$

$$\theta_{ecc} = 2\pi(t-1)/N \tag{13-3}$$

In dynamic eccentricity, the center of the rotor shifts by $e_d$, the magnitude of the eccentricity in millimeters, which rotates with the rotor at an angular velocity of $\omega$ in rad/s and causes a time-varying change in the airgap, as shown in (14-1). In this context, dynamic eccentricity can be integrated into the calculations of the terms $[L_{ee}(\theta) \quad L_{se}(\theta) \quad L_{ce}(\theta)]^T$ as a scaling factor, $f_{de}(\phi)$, shown in (14-2), where, $t$ denotes time.

$$g_{dynamic}(\phi, t) = g(\phi) + e_d.\cos(\varphi - \omega t) \tag{14-1}$$

$$f_{de}(\phi) = g(\phi)/(g(\phi) + e_d.\cos(\varphi - \omega t)) \tag{14-2}$$

The divisor $(G_{min} + G_{max}) + (G_{min} - G_{max})\cos\left(\frac{P}{2}\varphi\right)$ in (12-1) has contributions from $G_{max}$, $G_{min}$, and the cosine term (oscillates between $-1$ and 1). When $\cos\left(\frac{P}{2}\varphi\right) = \pm 1$, the change boundary of the airgap $g(\phi)$ is between $\frac{G_{min}}{2}$ and $\frac{G_{max}}{2}$. In (13-2) and (14-2), both $\cos(\varphi - \theta_{ecc})$ and $\cos(\varphi - \omega t)$ oscillate between $-1$ and 1; therefore, the effective airgap with static and dynamic eccentricity faults is written as $g(\phi) \pm e_s$ and $g(\phi) \pm e_d$, respectively. By substituting the lower bound of $g(\phi)$, $\frac{G_{min}}{2}$, and keeping in mind the physical feasibility, we can define the maximum eccentricity magnitudes as $\frac{G_{min}}{2}$. The lower boundary is zero, which is synonymous with the fault-free condition. In general, (6)-(8) can be rewritten, as shown in (15), by incorporating the scaling factors for static and dynamic eccentricity faults.

$$M_{se}[i,j] = T_{s_i} \cdot T_{e_j} \cdot L_{s_i e_j} \cdot f(\phi)$$
$$M_{ce}[i,j] = T_{c_i} \cdot T_{e_j} \cdot L_{c_i e_j} \cdot f(\phi) \quad (15\text{-}1)$$

$$l_{e_i e_j} = T_{e_i} \cdot T_{e_j} \cdot L_{e_i e_j} \cdot f(\phi)$$
$$l_{e_i e_i} = T^2{}_{e_j} L_{e_i e_i} \cdot f(\phi) \quad (15\text{-}2)$$

Fig. A1 is shown in the APPENDIX to summarize the suggested methodology at a glance through the written link.

## IV. Examination of Proficiency and Generality of the Hybrid Model

In the following, a 1-X resolver with a 12-slot stator, as shown in Fig. 1, is the case study. Owing to its non-uniform airgap and the asymmetric design of its rotor, it is a good enough candidate for the proficiency examination of the suggested hybrid model in the context of different winding configurations, including non-overlapping and overlapping configurations, as well as faulty conditions such as inter-turn short circuits in signal and excitation windings and static and dynamic eccentricity faults. The geometrical details of the investigated resolver are given in Table I. First and foremost, through a 2D TS-FEA, we calculate two matrices of size 144 for two groups of coils: excitation and signal, for the investigated resolver, while its stator slots contain single-turn coils with no series connection; the number of slots for signal and excitation windings is the same. Subsequently, we have the capability, along with wide flexibility, to calculate (6-1), (7-1), and (8-1). In Table II, diagrams of the investigated winding configurations are shown. In the non-overlapping configuration, each slot of the stator contains only one type of winding: either excitation or sine or cosine, with constant turns for all slots of a winding, whether signals or excitation. On the other hand, in the overlapping configuration, (16) gives the turn number and current direction of each coil at the $i$-th tooth of the excitation and signal windings [17] and [18].

$$T_{sin}(i) = T_s \sin\big(2\pi\, P_w (i - 1/N)\big) \quad (16\text{-}1)$$

$$T_{cos}(i) = T_s \cos\big(2\pi\, P_w (i - 1/N)\big) \quad (16\text{-}2)$$

$$P_w = P \text{ or } P_w = (N \pm 2P)/2 \quad (16\text{-}3)$$

$$T_{exc}(i) = \begin{matrix} T_e & \text{if } \mathrm{mod}(i,2) = 1 \\ -T_e & \text{if } \mathrm{mod}(i,2) = 0 \end{matrix} \quad (16\text{-}4)$$

where $T_s$ and $T_e$ are the maximum number of turns for the signal and excitation windings, respectively. Here, $T_s = 70$ and $T_e = 30$. Following results (2)-(5), the fault-free signal voltages are shown in Fig. 2 for both the hybrid model and 2D TS-FEA. Moreover, in Fig. 3, the terms of $[L_{ee}(\theta) \quad L_{se}(\theta) \quad L_{ce}(\theta)]^T$ resulting from the generalized hybrid model and 2D TS-FEA are displayed. Fig. 4 also shows the rotor angle errors for both analyses. For the sake of brevity, we study the faulty conditions only for the overlapping configuration. In this regard, a short circuit was injected into 21 turns out of 61 at the 9th tooth of the sine winding. Additionally, for the excitation winding short circuit, the 1st tooth was under fault with an intensity of 5 turns out of 30, with $R_{sc} = 0.2\ \Omega$ and $R_e = 2\ \Omega$. For both static and dynamic eccentricity faults, with $G_{min} = 0.5$ mm, the highest feasible intensity is 0.25 mm, while the investigated magnitude is 0.1 mm. In the static case, the stator shifts toward the 7th tooth with angular direction of $\theta_{ecc} = \pi$. Details of the calculations of matrices for the investigated resolver and faults are available in the APPENDIX link. In Table III, it can be seen that the suggested hybrid model, regarding AAPE and MPE metrics, has satisfactory accuracy in modeling the investigated resolver, with a worst-case error of (4.99%)⁄(4.73%) and (10.55%)⁄(12.61%) in healthy and faulty conditions, respectively, set against the 2D TS-FEA. This close agreement is also satisfactorily met in terms of building $L_{se}(\theta)$ and $L_{ce}(\theta)$ on the subject of Mean Absolute Error (MAE), as shown in Table IV. Moreover, since the magnitude of the excitation current is constant, $L_{ee}(\theta)$ remains almost constant with rotor rotation, as displayed in Fig. 3; therefore, its average value serves as the judgment metric for similarity [13].

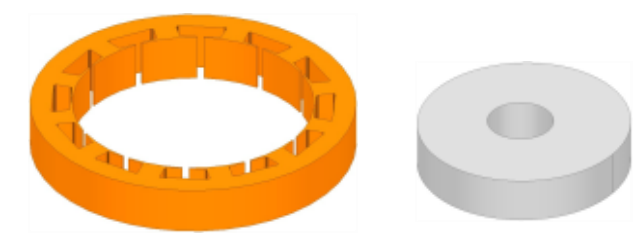

Fig. 1. 3D geometry of the investigated resolver: (a) 12-slot stator and (b) 1-X rotor.

TABLE I
GEOMETRICAL DETAILS OF THE INVESTIGATED RESOLVER

| | Parameter | Value | Unit |
|---|---|---|---|
| | Core Stack Length | 6.7 | mm |
| Stator | Outer/Inner Diameter | 45.96/34.13 | mm |
| | Slot Opening Height/Width | 0.2/0.99 | mm |
| | Number of Slots/Teeth | 12 | - |
| Rotor | Shaft Diameter | 10 | mm |
| | Minimum Airgap | 0.5 | mm |
| | Maximum Airgap | 2 | mm |
| | Pole Numbers ($P$) | 2 | - |
| | Number of Winding Pole Pairs ($P_w$) | 5 | - |

TABLE II
TWO WINDING DIAGRAMS OF THE INVESTIGATED RESOLVER

| | Non-Overlapping | | | Overlapping | | |
|---|---|---|---|---|---|---|
| Winding / Tooth | $W_s$ | $W_c$ | $W_e$ | $W_s$ | $W_c$ | $W_e$ |
| $N_1$ | 0 | 0 | $T_e$ ↷ | 0 | $T_s$ ↷ | $T_e$ ↷ |
| $N_2$ | $T_s$ ↷ | 0 | 0 | $0.50T_s$ ↷ | $0.87T_s$ ↶ | $T_e$ ↶ |
| $N_3$ | 0 | $T_s$ ↶ | 0 | $0.87T_s$ ↶ | $0.50T_s$ ↷ | $T_e$ ↷ |
| $N_4$ | 0 | 0 | $T_e$ ↷ | $T_s$ ↷ | 0 | $T_e$ ↶ |
| $N_5$ | 0 | $T_s$ ↷ | 0 | $0.87T_s$ ↶ | $0.50T_s$ ↶ | $T_e$ ↷ |
| $N_6$ | $T_s$ ↷ | 0 | 0 | $0.50T_s$ ↷ | $0.87T_s$ ↷ | $T_e$ ↶ |
| $N_7$ | 0 | 0 | $T_e$ ↷ | 0 | $T_s$ ↶ | $T_e$ ↷ |
| $N_8$ | $T_s$ ↶ | 0 | 0 | $0.50T_s$ ↶ | $0.87T_s$ ↷ | $T_e$ ↶ |
| $N_9$ | 0 | $T_s$ ↷ | 0 | $0.87T_s$ ↷ | $0.50T_s$ ↶ | $T_e$ ↷ |
| $N_{10}$ | 0 | 0 | $T_e$ ↷ | $T_s$ ↶ | 0 | $T_e$ ↶ |
| $N_{11}$ | 0 | $T_s$ ↶ | 0 | $0.87T_s$ ↷ | $0.50T_s$ ↷ | $T_e$ ↷ |
| $N_{12}$ | $T_s$ ↶ | 0 | 0 | $0.50T_s$ ↶ | $0.87T_s$ ↶ | $T_e$ ↶ |

$W_s$: sine winding, $W_c$: cosine winding, and $W_e$: excitation winding
The ↷ denotes the winding direction: ⊗↷⊙

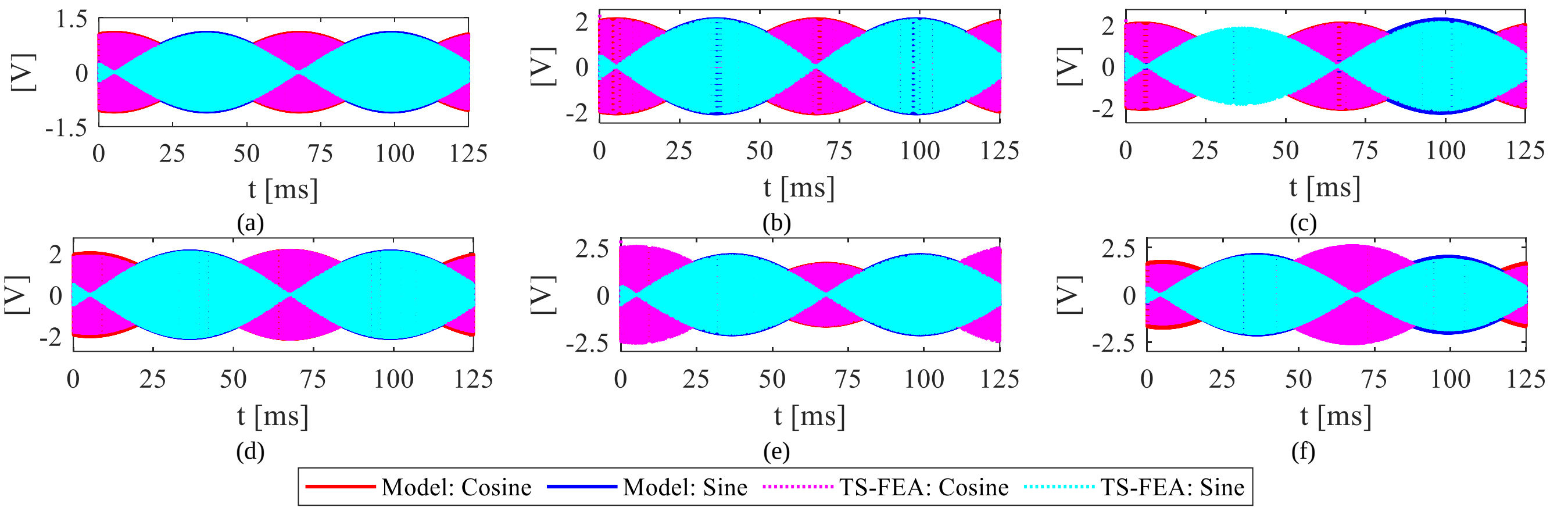

Fig. 2. Signal voltages of the generalized model and TS-FEA.

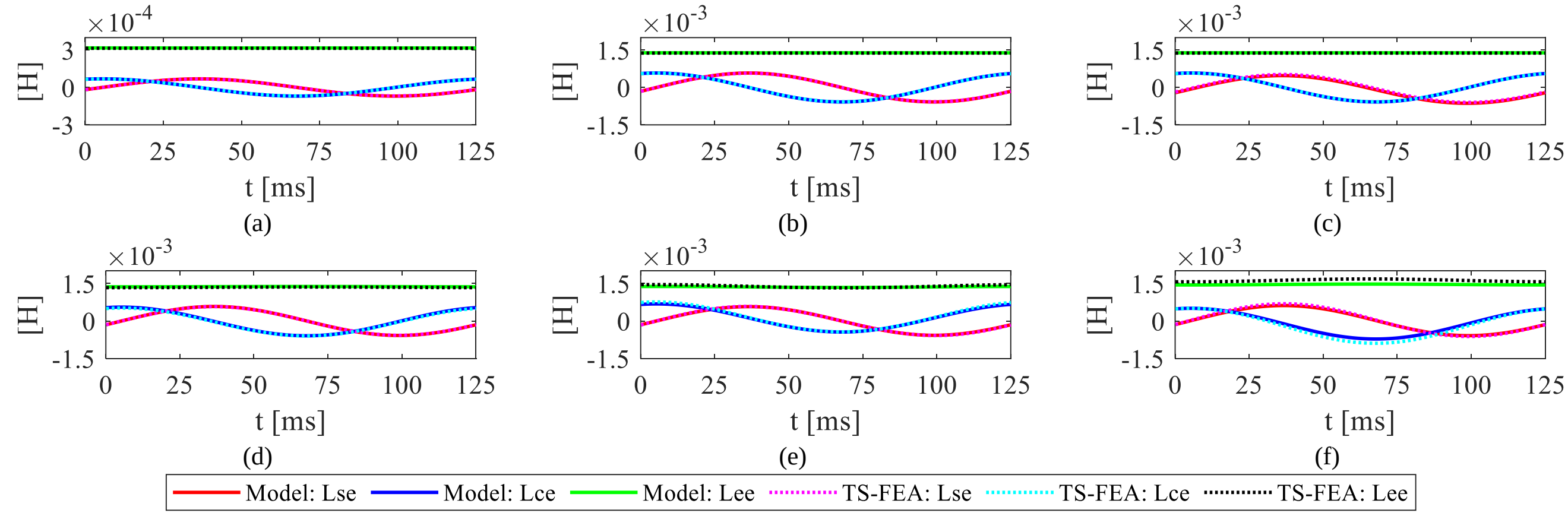

Fig. 3. The terms of $[L_{ee}(\theta) \quad L_{se}(\theta) \quad L_{ce}(\theta)]^T$ for the generalized model and TS-FEA.

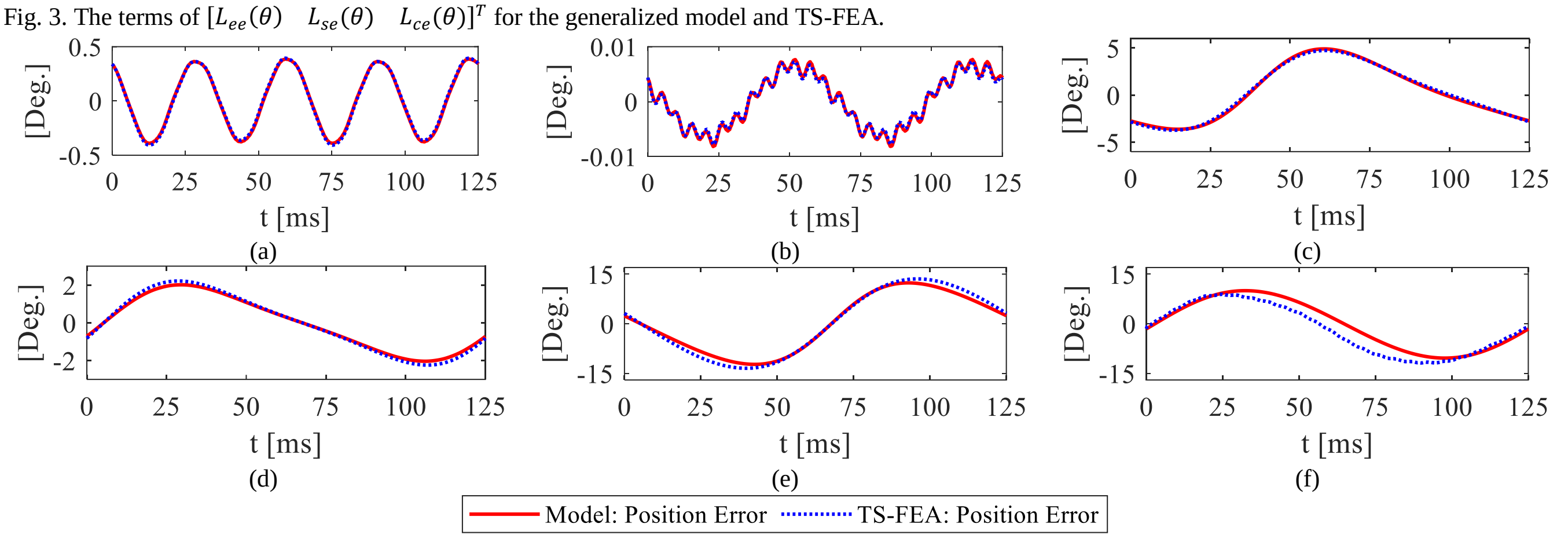

Fig. 4. Rotor angle errors for both the generalized model and TS-FEA. In Figs. 2, 3, and 4, the designations are as follows: (a) non-overlapping winding in a healthy operating condition; (b) overlapping winding in a healthy operating condition; (c) inter-turn short circuit at the 9th tooth of the sine winding; (d) inter-turn short circuit at the 1st tooth of the excitation winding; (e) static eccentricity under the 7th tooth; and (f) dynamic eccentricity.

TABLE III
COMPARISON OF RESULTS FROM THE HYBRID MODEL AND TS-FEM

| Winding Config | Operating Condition | AAPE (Deg.) | | | MPE (Deg.) | | | Time efficiency (min.) | | | | |
|---|---|---|---|---|---|---|---|---|---|---|---|---|
| | | TS-FEA | Model | Error (%) | TS-FEA | Model | Error (%) | TS-FEA | Model | | | Time saving (%) |
| | | | | | | | | | Initializing | Calculation | Total | |
| Non-Overlapping | Healthy | 0.24 | 0.24 | 0.91 | 0.41 | 0.39 | 4.12 | ≅ 250 | ≅ 25 | ≤ 1 | ≅ 26 | 89.60 |
| Overlapping | Healthy | 0.004 | 0.004 | 4.99 | 0.008 | 0.008 | 4.73 | ≅ 750 | | | | 96.53 |
| | Fault 1 | 2.60 | 2.62 | 0.55 | 4.73 | 4.90 | 3.61 | | | | | |
| | Fault 2 | 1.25 | 1.14 | 8.62 | 2.24 | 2.04 | 9.10 | | | | | |
| | Fault 3 | 7.99 | 7.15 | 10.55 | 13.56 | 12.35 | 8.90 | | | | | |
| | Fault 4 | 6.36 | 6.09 | 4.29 | 11.88 | 10.38 | 12.61 | | | | | |

Fault designations are as follows: (1) signal winding inter-turn short circuit, (2) excitation winding inter-turn short circuit, (3) static eccentricity, and (4) dynamic eccentricity.

TABLE IV
HYBRID MODEL AND TS-FEM INDUCTANCE COMPARISON

| Winding Config. | Operating Condition | MAE ($\mu H$) | | Avg. $L_{ee}$ ($mH$) | | |
|---|---|---|---|---|---|---|
| | | $L_{se}$ | $L_{ce}$ | FEA | Model | Error (%) |
| Non-Overlapping | Healthy | 0.7 | 0.8 | 0.32 | 0.32 | 0.8 |
| Overlapping | Healthy | 3.5 | 3.9 | 1.38 | 1.39 | 0.8 |
| | Fault 1 | 38.8 | 3.9 | 1.38 | 1.39 | 0.8 |
| | Fault 2 | 3.4 | 23.1 | 1.35 | 1.37 | 2.0 |
| | Fault 3 | 7.8 | 34.8 | 1.40 | 1.36 | 2.3 |
| | Fault 4 | 34.1 | 78.1 | 1.62 | 1.46 | 11.1 |

Fault designations are the same as in Table III.

$$\text{MAE} = \frac{1}{t}\sum_{i=0}^{t}|y_{FEA}(i) - y_{Model}(i)| \quad (17)$$

The results of the faulty conditions for both the hybrid model and 2D TS-FEA, including signal voltages, the terms $[L_{ee}(\theta) \quad L_{se}(\theta) \quad L_{ce}(\theta)]^T$, and rotor angle errors, are displayed in Figs. 2, 3, and 4.

If we want to maintain the accuracy of the resolver while using a TS-FEA to extract the resolver signal voltages, a sampling frequency at least 16 times greater than its excitation frequency must be chosen, which is synonymous with extremely low time efficiency [19]. In this case, such analysis takes, on average, 480 minutes for different windings with a 5 kHz excitation frequency and an 80 kHz sampling frequency (Intel® Core™ i5-7400 CPU @ 3.00GHz). On the other hand, following the Nyquist frequency rule [20], we can select a lower sampling frequency for the 2D TS-FEA of the generalized hybrid model. Here, the resolver rotates at a frequency of 8 Hz (each rotation takes 125 ms), which amounts to an angular velocity of 480 r/min ($\omega$ = 50.27 rad/s). Hence, to account for the 500th frequency order of the 8 Hz frequency within the FFT, we should use at least an 8 kHz sampling frequency, which is 10 times lower than that deployed in the full TS-FEA, and the generalized model ensures lower complexity in the windings. In this case, such analysis takes almost 25 minutes. Here, the number of mesh elements is 102574. After the TS-FEA, the suggested model can be easily managed in MATLAB/SIMULINK, taking less than 1 minute for each analysis. The effect of such time savings will turn into an increasingly significant factor when we want to study different types of windings and faulty conditions, especially varying degrees of fault intensity and fault overlapping; the time taken for the generalized hybrid model grows with each further analysis for the TS-FEA. In other words, the generalized hybrid model under healthy conditions is 70% faster than its first generation, according to the results in [14], and approximately 96% faster than TS-FEA. In investigations of faulty conditions, the time taken by the generalized hybrid model remains constant for an F number of faulty conditions and winding configurations. In contrast, for the same number of analyses, the time taken by the first-generation hybrid model and TS-FEA can be written as $\text{F} \times \text{time}$, which grows exponentially. Table A1 in the APPENDIX link shows the suggested methodology's proficiency against MEC and WF methodologies.

## V. EXPERIMENTAL MEASUREMENTS

Finally, we built the investigated 1-X 12-slot resolver, as shown in Fig. 5, with an overlapping winding configuration to evaluate the level of agreement between the results from the generalized hybrid model and real-world measurements. In this regard, the test bench shown in Fig. 6 was gathered. A DC motor serves as the primary motor, with coupling to the prototyped resolver and an 18-bit absolute encoder. The angular velocity of the motor is adjustable via its voltage supply. Here, the encoder acts as the reference sensor. The resolver's resulting signal voltages, displayed in Fig. 7 for both healthy and faulty operating conditions, were analyzed using MATLAB software to bypass the corrective effect of the Resolver-to-Digital Converter (RDC), leading to the calculation of the error between the resolver rotor angle and the encoder angle. Due to the challenges of manufacturing resolvers with certain faults and the severe damage that these faults—such as dynamic eccentricity and short circuits—could cause to the case study, this test focuses solely on static eccentricity by using a damaged asymmetric stator frame with different centers for its interior and exterior circles. An automatic gain control unit adjusts the excitation voltage, while a digitally synthesized function generator feeds the excitation winding in the resolver circuit. The excitation voltage is a 5 V sine wave at 5 kHz (with 0.1 Hz resolution). The laboratory conditions are categorized as "grounded and stationary," indicating a lack of mobility, and the ambient heat and humidity are under control [21]. The evaluation metrics, AAPE and MPE, show errors of $5.09\%/4.96\%$ and $11.38\%/10.89\%$, respectively, under fault-free conditions and for static eccentricity (0.1 mm magnitude under the 7th stator tooth). This shows a satisfactory agreement for both healthy and faulty conditions between the generalized hybrid model and real-world measurements.

## VI. CONCLUSION

In this study, a generalized hybrid resolver model has been suggested to not only save significant time relative to other methodologies, such as TS-FEA, with close agreement in accuracy to them, but also reduce the number of initializing analyses to the least, due to its proficiency and generality in modeling different winding configurations and faulty conditions: here, inter-turn short circuit and eccentricity, just by a one-time calculation of basis functions for single-turn coils within stator/rotor slots. In this regard, the methodology of the generalized model has been shown in detail, including how to integrate different faulty conditions. Then, it has been shown that the suggested model has satisfactory agreement with TS-FEA in both healthy and faulty conditions, with, in the worst case, about 5% and 13% differences in AAPE and MPE metrics, respectively, while saving at least 96% of the time, which behaves exponentially when conducting a reliability study. At the end, to verify the suggested model experimentally, the case study was prototyped and went under lab test, showing close agreement in results between the suggested model and real-world measurements. It is worth mentioning that even with the merits of the suggested methodology, it has shortcomings for future work: (1) The hybrid model could address nonlinear B-H curve effects, particularly for resolvers under extreme operating conditions or high saturation. (2) Other faults (e.g., rotor PM demagnetization and runout) could be integrated to strengthen the methodology's applicability.

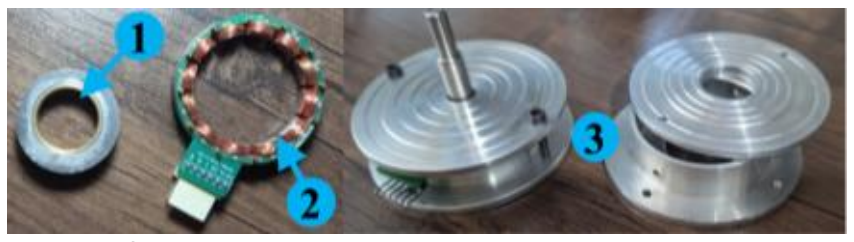

Fig. 5. Prototype of a 12-slot 1-X resolver with an overlapping winding configuration: (1) the 1-X rotor, (2) the wound stator and (3) built frames (healthy on the left and faulty on the right – with a static eccentricity of 0.1 mm under the 7th tooth of the stator).

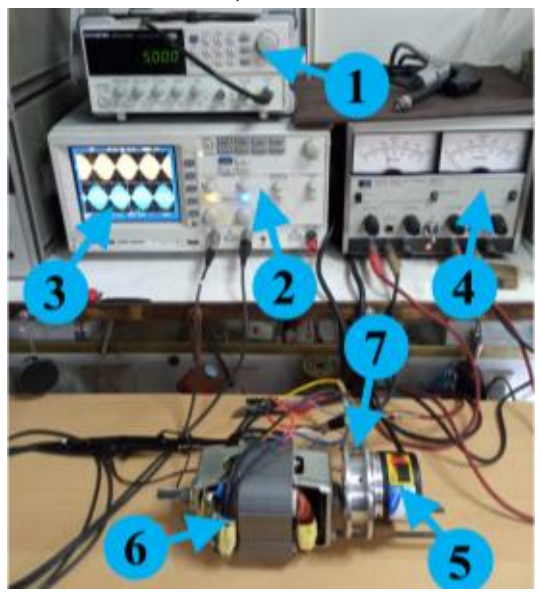

Fig. 6. Test bench, including (1) Function Generator, (2) Digital Oscilloscope, (3) Signal Voltages, (4) DC Power Supply, (5) Optical Encoder, (6) DC Motor, and (7) Built Resolver.

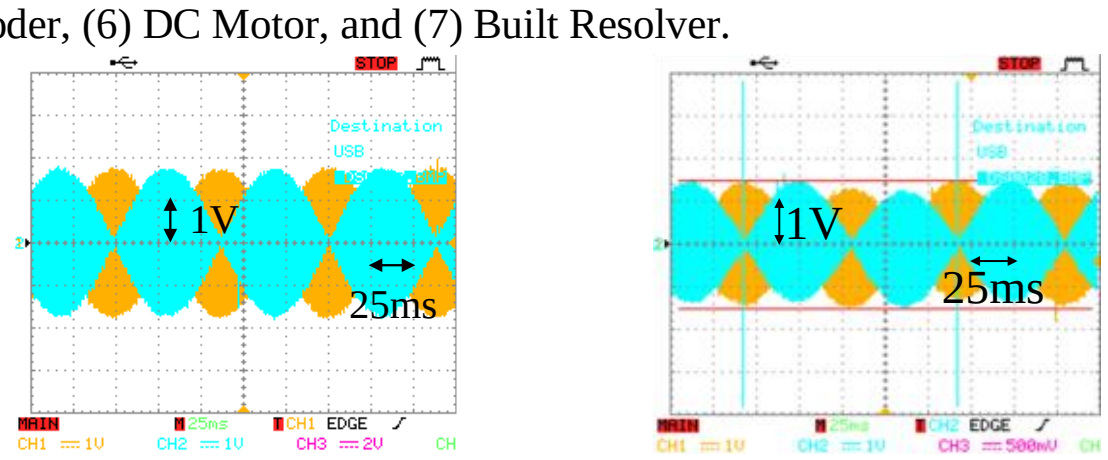

(a) (b)

Fig. 7. Comparison of the signal voltages from the prototyped resolver in both (a) healthy and (b) faulty conditions (with 0.1 mm static eccentricity).

## APPENDIX

Supplementary data are available in the Google Drive link to offer extra clarity and meet the page limit:

https://drive.google.com/file/d/1ZgacuqPB3SdqlaIOf7BoDGEBOw8pYjPH/view?usp=sharing